\documentstyle[12pt]{article}
\begin{document}
\title{\large \hspace{10cm} ITEP-44/99 \\ \hspace{10cm} December 1999 \\
\vspace{1cm}
\LARGE \bf
Non-BPS domain wall configurations in a supersymmetric model}
\author {V. A. GANI\thanks{E-mail: gani@heron.itep.ru}{\,}
\\
{\it Moscow State Engineering Physics Institute (Technical University),}\\
{\it Kashirskoe shosse, 31, Moscow, 115409, Russia}\\
{\it and}\\
{\it Institute of Theoretical and Experimental Physics, Russia}\\
\\
A. E. KUDRYAVTSEV \thanks{E-mail: kudryavtsev@vitep5.itep.ru}
\\
{\it Institute of Theoretical and Experimental Physics,}\\
{\it B.Cheremushkinskaya, 25, Moscow, 117259, Russia}\\
}
\date{}
\maketitle
\vspace{1mm}
\centerline{\bf {Abstract}}
\vspace{3mm}
     We study the time evolution of configurations in the form of two
parallel domain walls moving towards each other in a supersymmetric
field model. The configurations involved are not BPS-saturated.
It is found that for such collisions there exists some critical value
$v_{cr}\approx0.9120$ of the initial velocity $v_i$ of the
walls. At $v_i<v_{cr}$ we observed reflection, that was not followed by
change of vacuum states sequence. In collisions with $v_i>v_{cr}$
the sequence of vacuum states changes. The results of the numerical
simulations are in agreement with "potential" consideration.

\newpage

     The dynamic properties of domain walls in supersymmetric theories
have attracted some attention recently \cite{ShV}~-~\cite{GK}.
Depending on particular form of the superpotential being chosen,
one obtains different sets of supersymmetric vacua and different
structure of domain wall configurations interpolating between them.
We restrict ourselves by consideration of the theory described by
the superpotential
$$
W(\Phi,X)=\frac{m^2}{\lambda}\Phi-\frac{1}{3}\lambda\Phi^3-\alpha\Phi X^2 \ ,
\eqno(1)
$$
where $m$ is a mass parameter and $\alpha$ and $\lambda$ are coupling
constants. We assume that $\alpha$ and $\lambda$ are real and positive.
The Lagrangian for the real parts of the scalar fields is given for this
theory by the expression
$$
L=(\partial\phi)^2+(\partial\chi)^2
-\left(\frac{m^2}{\lambda}-\lambda\phi^2-\alpha\chi^2\right)^2
-4\alpha^2\phi^2\chi^2 \ .
\eqno(2)
$$
The potential term of Eq.~(2) has four degenerate vacuum states, shown in
Fig.~1. This theory possess a wide class of domain walls, which link different
vacua. Some of them satisfy first order differential equations analogous to
the Bogomol'nyi-Prasad-Sommerfeld (BPS) equations~\cite{BPS}. The dynamic
properties of the BPS configurations for this model were intensively studied
recently \cite{ShV}~-~\cite{GK}. This our work is devoted to the so-called
non-BPS domain walls, i.e. configurations which link different vacua of
the theory, but do not satisfy BPS equations.

     It is convenient to work with dimensionless field variables $f$ and $h$,
defined as
$$
\phi=\frac{m}{\lambda}f \ , \ \ \chi=\frac{m}{\sqrt{\lambda\alpha}}h \ .
$$
The Lagrangian (2) yields the following equations of motion for fields
$f$ and $h$:
$$
f_{tt}^2-\nabla^2f-2f(1-f^2-h^2)+\frac{4}{\rho}fh^2=0 \ ,
$$
$$
h_{tt}^2-\nabla^2h-\frac{2}{\rho}h(1-f^2-h^2)+\frac{4}{\rho^2}hf^2=0 \ .
\eqno(3)
$$
Here $\rho=\lambda/\alpha$, $m=1$. It was shown
(see, e.g., Ref.~\cite{TV}) that for the case $\rho=4$ the field
equations (3) possess an "elementary" walls, connecting vacua
$3$ and $2$, $2$ and $4$. Their form may be obtained
analytically~\cite{TV}:
$$
f_{32}(z)=\frac{1}{2}\left(1+\tanh{\frac{z}{2}}\right) , \
h_{32}(z)=\sqrt{\frac{1}{2}\left(1-\tanh{\frac{z}{2}}\right)} \ ;
\eqno(4)
$$
$$
f_{24}(z)=\frac{1}{2}\left(1-\tanh{\frac{z}{2}}\right) , \
h_{24}(z)=-\sqrt{\frac{1}{2}\left(1+\tanh{\frac{z}{2}}\right)} \ ,
\eqno(5)
$$
here $z$ is a space coordinate orthogonal to the walls. It is easy
to see, that the rest energy of these $3\to2$ and $2\to4$ walls
equals $E_0=4/3$.

     Consider a non-BPS ansatz configuration $3\to2\to4$ constructed from
two elementary domain walls $3\to2$ and $2\to4$ located at $z=-z_0$
and $z=+z_0$ respectively. Let us take their simple superposition
in the form
$$
f_{324}(z,z_0)=f_{32}(z+z_0)+f_{24}(z-z_0)-1 \ ,
$$
$$
h_{324}(z,z_0)=h_{32}(z+z_0)+h_{24}(z-z_0) \ .
\eqno(6)
$$
Note, that from the system (3) the special "diagonal" solution $3\to4$
can be easily found by substituting $f=0$:
$$
f_{34}(z)\equiv0 , \ h_{34}(z)=-\tanh{\frac{z}{2}} \ .
\eqno(7)
$$
The energy of such configuration is $E_{34}=16/3$.

     To get $z_0$-dependence of the energy of configuration (6) we
have to insert (6) into Hamiltonian of the model. As a result we obtain
$$
E_{324}(z_0)=2E_0+\Delta E_{324}(z_0) \ ,
\eqno(8)
$$
where
$$
\Delta E_{324}=\int\limits_{-\infty}^{+\infty}dz\left[2\frac{df_{32}}{dz}
\frac{df_{24}}{dz}+2\rho\frac{dh_{32}}{dz}\frac{dh_{24}}{dz}
+(1-f_{324}^2-h_{324}^2)^2
+\frac{4}{\rho}f_{324}^2h_{324}^2
\right.
$$
$$
\left.
-(1-f_{32}^2-h_{32}^2)^2-\frac{4}{\rho}f_{32}^2h_{32}^2
-(1-f_{24}^2-h_{24}^2)^2-\frac{4}{\rho}f_{24}^2h_{24}^2
\right] \ .
\eqno(9)
$$
Here $f_{32}=f_{32}(z+z_0)$, $h_{32}=h_{32}(z+z_0)$,
$f_{24}=f_{24}(z-z_0)$, $h_{24}=h_{24}(z-z_0)$.
We calculated the $z_0$-dependence of $\Delta E_{324}$ numerically,
see Fig.~2 (solid curve). At the limit of large $z_0$
the configuration (6) looks like two isolated walls
$3\to2$ and $2\to4$. Therefore their total energy equals $2E_0$,
and $\Delta E_{324}\approx0$. As it is seen from Fig.~2, energy
$\Delta E_{324}$ increases with decreasing $z_0$.
At $z_0=0$ $\Delta E_{324}(0)\approx3.119$. It corresponds to
$E_{324}(0)=2E_0+\Delta E_{324}(0)\approx 5.786$.
Note, that $E_{324}(0)$ is larger than $E_{34}=16/3\approx5.333$.
The energy of configuration $3\to2\to4$ (6) has its absolute
maximum at $z_0\approx-0.37$ when $(\Delta E_{324})_{max}\approx3.202$.
At large negative $z_0$ $\Delta E_{324}(z_0)$ has asymptotic value
about $2.274$. In the range $z_0<0$ configuration (6) actually
has the shape of the $3\to1\to4$ type, see Fig.~3.

     It is clear, that we can construct an ansatz configuration
$3\to1\to4$ in analogy to (6):
$$
f_{314}(z,z_0)=f_{31}(z+z_0)+f_{14}(z-z_0)+1 \ ,
$$
$$
h_{314}(z,z_0)=h_{31}(z+z_0)+h_{14}(z-z_0) \ ,
\eqno(10)
$$
where
$$
f_{31}(z)=-\frac{1}{2}\left(1+\tanh{\frac{z}{2}}\right) , \
h_{31}(z)=\sqrt{\frac{1}{2}\left(1-\tanh{\frac{z}{2}}\right)} \ ;
\eqno(11)
$$
$$
f_{14}(z)=-\frac{1}{2}\left(1-\tanh{\frac{z}{2}}\right) , \
h_{14}(z)=-\sqrt{\frac{1}{2}\left(1+\tanh{\frac{z}{2}}\right)} \ .
\eqno(12)
$$
The energy of $3\to1$ and $1\to4$ walls is exactly the same as
of $3\to2$ or $2\to4$. Hence, the $z_0$-dependence of the energy
of configuration (10) will be
$$
E_{314}(z_0)=2E_0+\Delta E_{314}(z_0) \ ,
\eqno(13)
$$
with "potential" $\Delta E_{314}(z_0)$ analogous to
$\Delta E_{324}(z_0)$, Eq.~(9).
The shape of $\Delta E_{314}(z_0)$ is the same as of
$\Delta E_{324}(z_0)$.
As it was already mentioned, at $z_0<0$ ansatz (6)
has the shape of $3\to1\to4$ type indeed. Obviously,
configuration (10) at negative $z_0$ has the shape of
$3\to2\to4$ type, see Fig.~3.
Notice, that $h_{324}(z,-z_0)\ne h_{314}(z,z_0)$.
If we would like to compare energies of configurations (6) or (10)
that belong to one of these two types, we have to place
curves $\Delta E_{324}(z_0)$ and $\Delta E_{314}(-z_0)$
(or curves $\Delta E_{314}(z_0)$ and $\Delta E_{324}(-z_0)$)
in the same plot. Fig.~2 is constructed just in this way.

     We solved field equations (3) numerically with
initial conditions in the form of (6), where $3\to2$ and $2\to4$
walls located at some initial distance $2z_0\gg1$ and are moving
towards each other with some initial velocity $v_i$.
Depending on the initial velocity we observed different types of
evolution. If $v_i$ is less than some critical value $v_{cr}^{num}$,
walls $3\to2$ and $2\to4$ collide and then escape from each other
to infinity. As a result we return to the configuration of
the $3\to2\to4$ type. At initial velocities $v_i>v_{cr}^{num}$
the walls collide in a different way. The point is that after
collision the configuration of the type $3\to1\to4$ appears.
From these numerical simulations we found $v_{cr}^{num}\approx0.9120$.
The presence of different regimes in such collisions is
a consequence of the fact that the energy of configuration
$3\to2\to4$ is not degenerate with respect to the parameter $z_0$.
So, we have a kind of "potential" interaction between $3\to2$ and
$2\to4$ domain walls. It is worth to mention here, that in the case
of BPS-saturated (or simply BPS) walls~\cite{GK} there is no potential
interaction. The latter property is a consequence
of the degeneracy in energies of configurations with different
interwall distances, analogous to our parameter $z_0$.

     Existence of the critical velocity can be understood in terms of
the potential approach. From Fig.~2 it is seen, that if the initial
kinetic energy of the walls $3\to2$ and $2\to4$ is smaller than
$\Delta E^*\approx3.119$, then (inelastic) reflection may be expected.
If the kinetic energy of the walls exceeds $\Delta E^*$, it is natural
to expect that configurations of the type (6) with negative $z_0$
appear. But configuration (6) at negative $z_0$ is of the type
$3\to1\to4$ indeed, and from Fig.~2 we see, that in this sector
configurations (10) have smaller energy. Hence, configuration (6)
at negative $z_0$ transforms into (10). In further evolution the walls
$3\to1$ and $1\to4$ escape to infinity.

     It is also worth mentioning, that the initial configuration (6)
with $z_0=0$ and $v_i=0$ looks like some excitation over the static
solution (7). After emission of part of energy in the form of waves,
the evolution of this initial configuration ($z_0=0$, $v_i=0$) leads
to formation of an excited kink of type (7) (wobbling kink). We were
unable to get this wobbling kink solution making numerical calculations
of the equations of motion (3) with initial conditions (6) when
either $z_0$ or $v_i$ was not equal to zero.

\begin{center}
\bf
Acknowledgments
\end{center}

     We are thankful to M.~B.~Voloshin for useful discussions.
One of the authors (V.~A.~Gani) would like to thank E.~A.~Smirnova
for placing in our disposal some hardware resources and data transfer
channel.

     This work was supported in part by the Russian Foundation for Basic
Research under grants No~98-02-17316 and No~96-15-96578.
The work of V.~A.~Gani was also supported by the INTAS Grant No~96-0457
within the research program of the International Center for Fundamental
Physics in Moscow.

\newpage

\begin{center}
\bf
Figure captions
\end{center}
\bigskip

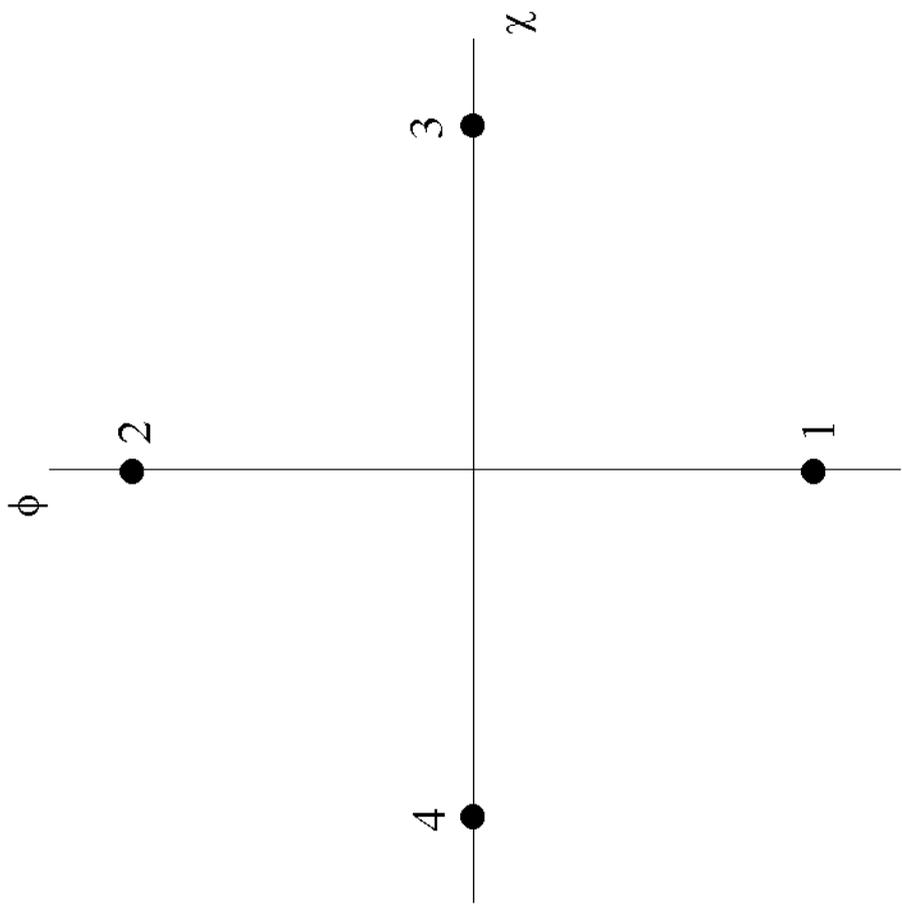
{\bf Fig.~1.} Locations of the vacuum states of the model.

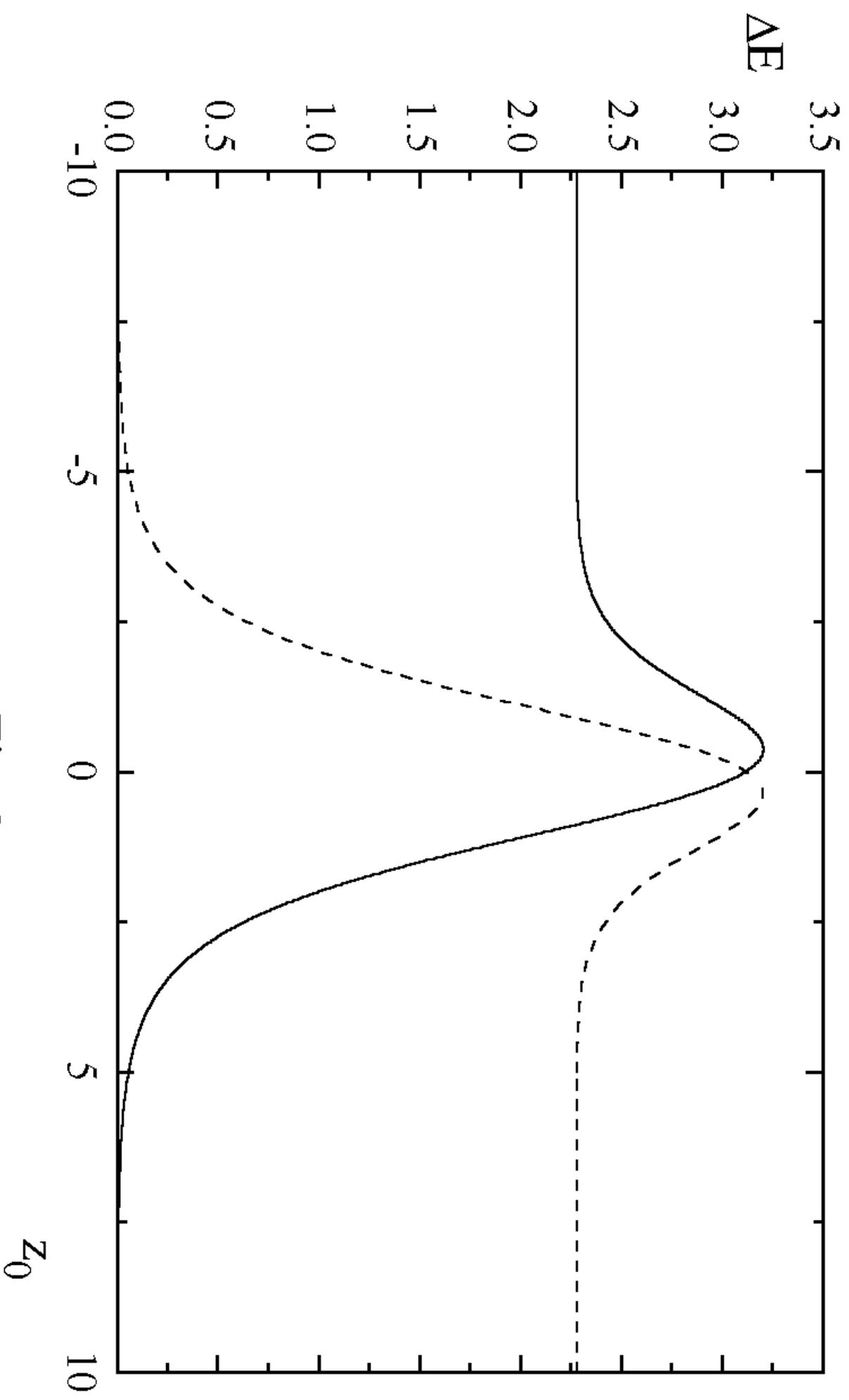
{\bf Fig.~2.} The profile of the potential $\Delta E_{324}$
              versus $z_0$ (solid curve)
              and the profile of the potential $\Delta E_{314}$
              versus $(-z_0)$ (dashed curve).

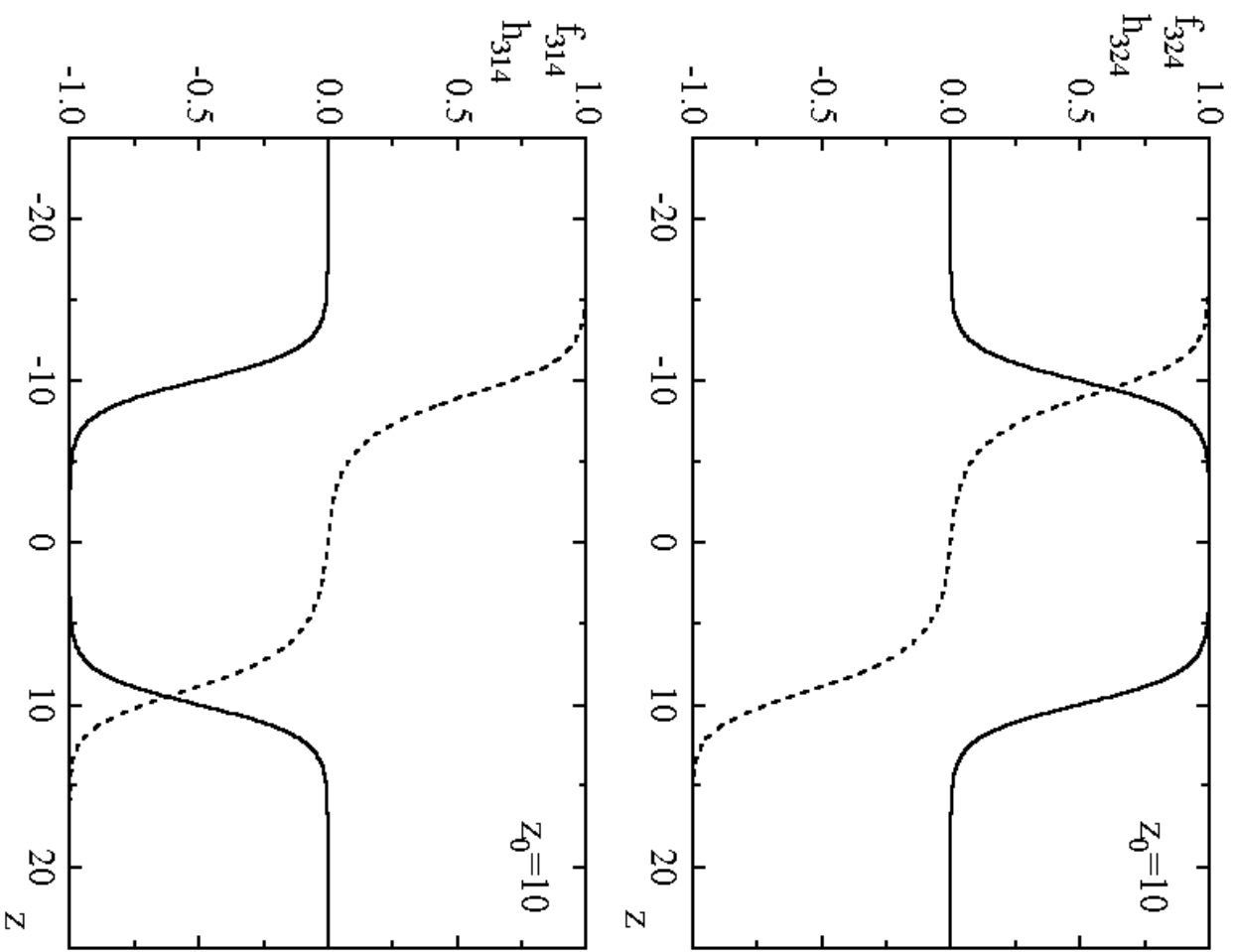
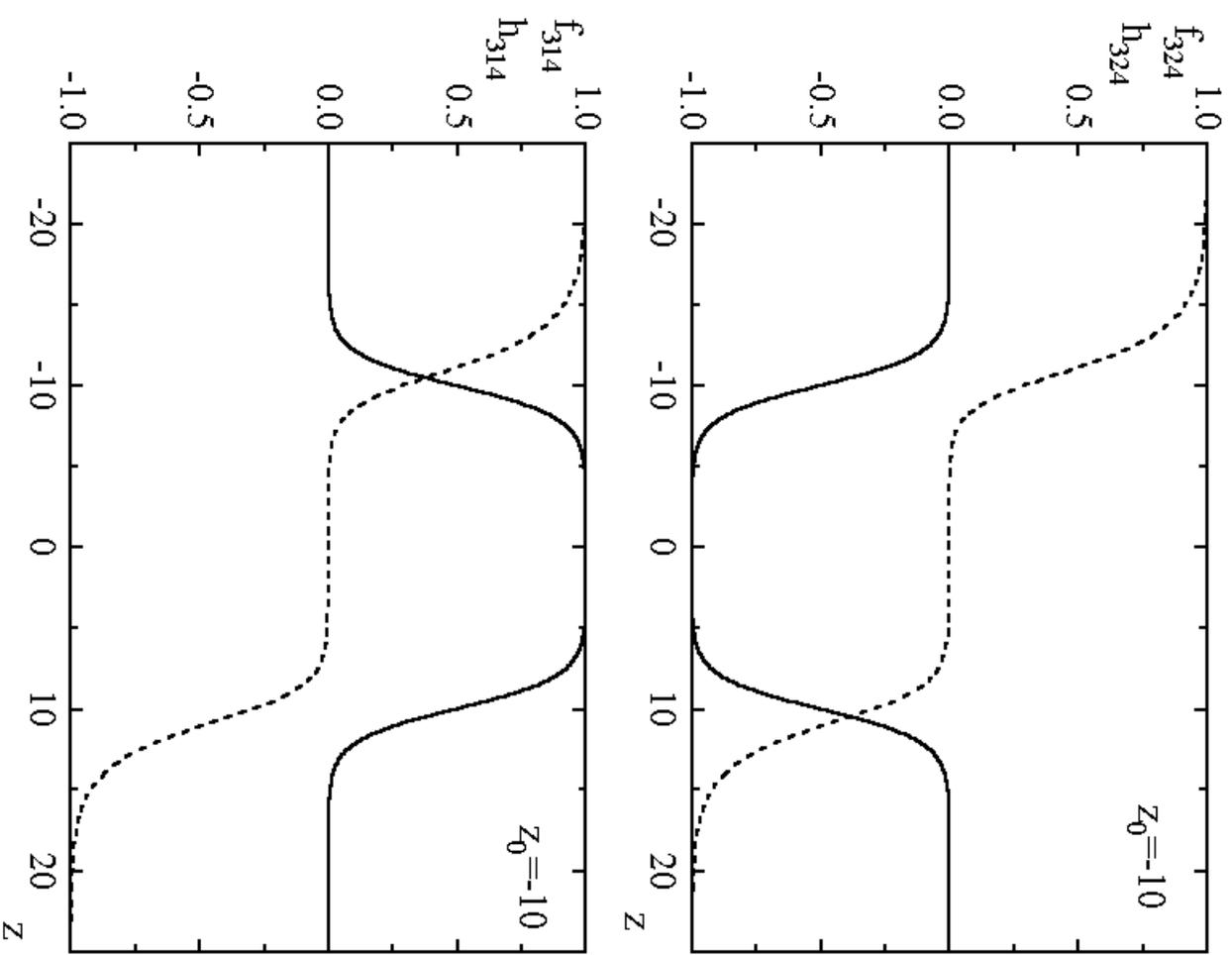
{\bf Fig.~3.} Profiles of $f(z)$ (solid lines) and $h(z)$ (dashed lines)
              for configurations $3\to2\to4$ and $3\to1\to4$
              at $z_0=\pm10.0$.

\end{document}